\documentstyle[preprint,aps,epsf]{revtex}

\newcommand{\bi}{\bibitem}
\newcommand{\be}{\begin{eqnarray}}
\newcommand{\ee}{\end{eqnarray}}
\newcommand{\nn}{\nonumber}
\catcode`\@=11
\def\lsim{\mathrel{\mathpalette\@versim<}}
\def\gsim{\mathrel{\mathpalette\@versim>}}
\def\@versim#1#2{\vcenter{\offinterlineskip
\ialign{$\m@th#1\hfil##\hfil$\crcr#2\crcr\sim\crcr } }}
\catcode`\@=12

\begin{document}

\tightenlines

\draft
\preprint{KANAZAWA-02-29, KUNS-1810,  OSU-HEP-02-19, IFUNAM-FT2002-06}
\title{Finite Theories and the SUSY Flavor Problem}
\author{K. S. Babu}
\address{Department of Physics, Oklahoma State University,
Stilwater, OK 74078 USA}
\author{Tatsuo Kobayashi}
\address{Department of Physics, Kyoto University,
Kyoto 606-8502, Japan}
\author{Jisuke Kubo\footnote{
Permanent address:
Institute for Theoretical Physics,
Kanazawa  University,
Kanazawa 920-1192, Japan}}
\address{Instituto de F\' isica,  UNAM,
Apdo. Postal 20-364,
M\' exico 01000 D.F., M\' exico}
\maketitle
\begin{abstract}
We study a finite $SU(5)$ grand unified model based on the
non-Abelian discrete symmetry $A_4$.
This model leads to the democratic structure of the 
mass matrices for  the quarks and leptons.
In the soft supersymmetry breaking sector, the scalar trilinear 
couplings are aligned and the soft scalar masses are degenerate,
thus solving the SUSY flavor
problem.

\end{abstract}


\section{Introduction}
Supersymmetry (SUSY) is broken in nature.
It is widely accepted  that the effects of supersymmetry breaking appear
as soft supersymmetry breaking (SSB) terms \cite{susy}.
However, if only renormalizability is used to guide
the SSB parameters, it is possible to
introduce more than 100 new parameters into the
minimal supersymmetric standard model (MSSM) \cite{dimopoulos1}.
The problem is not only this large number of
the independent parameters, but also the fact that one has to
highly fine tune these parameters so that they do not cause
problems with experimental observations
on the flavor changing neutral current (FCNC) processes and CP-violation
phenomena \cite{fcnc-mueg,fcnc-k,fcnc-edm,fcnc-bsg,fcnc}.
This problem,
called the SUSY flavor problem, is not new, but has existed ever
since supersymetry found phenomenological 
applications \cite{dimopoulos2}.

There are several approaches to overcome this problem.
The most well-known one \cite{susy} is to simply assume that
the SSB parameters have a universal form, independent of the
flavor structure of the standard model (SM) at, say, GUT scale
$M_{\rm GUT}$.
This is the so-called minimal supergravity model.
In this model, supersymmetry breaking occurs in a
sector that is hidden to the MSSM sector, and supersymmetry breaking
is mediated to the MSSM sector by gravity.
There exist other ideas of mediation:
gauge mediation \cite{gauge}, anomaly mediation \cite{anomaly} and gaugino
mediation \cite{gaugino}.
Their common feature is the assumption
 that there exists a hidden sector that is separated from
the MSSM by  cleverly chosen
interactions or it is separated  in  space time (for which
one needs extra dimensions
\footnote{There is a problem
associated with this approach,
the problem of
sequestering of branes between the
visible sector and the hidden SSB sector \cite{string}.}), or both.
Another type of idea to overcome the SUSY flavor problem is to
use the infrared attractive force of
the gauge interactions \cite{ross}, in
four dimensions \cite{karch,ns,ns1,nkt,ls} as well as
in extra dimensions \cite{kubo6}.
In these scenarios, it is not necessary to assume that
the supersymmetry is broken in a sector that is separated from the MSSM.

Although it is attractive to find  dynamical mechanisms that
suppress the dangerous  FCNC processes and CP-violating
phases, it is also worthwhile to look for
other attractive possibilities.
In fact, it has been argued \cite{parkes1,jones1,jack3,kazakov3}
that finiteness  \footnote{See
\cite{PW,HPS1,HPS2,mode2} for earlier references on
finite theories. } of softly broken supersymmetric
Yang-Mills theories \cite{lucchesi1,kapetanakis,kazakov2,yoshioka1,kubo5}
may play an important role to
understand the universality of the SSB parameters.
However, it has turned out \cite{kubo5,kubo4}
that the universality is not a necessary condition
for finiteness: 
It has been found \cite{kubo5,kubo4} that more
relaxed conditions, sum rules among the soft scalar masses,
are sufficient. Clearly,
the sum rules do not  automatically
ensure  the  degeneracy of the soft scalar masses.
In fact, 
in the unified models of
\cite{kapetanakis,kazakov2,yoshioka1,kubo5}
in which the hierarchical structure of the Yukawa couplings
emerges, the sum rules
cannot sufficiently
constrain the individual soft scalar masses in the first two generations
of the squarks and sleptons.

Recently, a class of finite models
based on $SU(5)$ with certain discrete symmetries
has been considered in \cite{babu1}, and
it has been found that some of these models yield a democratic structure
of the Yukawa couplings.
As we will see, the democratic structure is essential to obtain
sum rules of the soft scalar masses from which
 their degeneracy in generation follows.
Therefore, the exact finiteness in the models of \cite{babu1}
ensures the absence of the SUSY flavor problem.

In Sect. II, we will recapitulate the finiteness
conditions in softly broken supersymmetric
Yang-Mills theories \cite{lucchesi1,kapetanakis,kazakov2,yoshioka1,kubo5}.
The unified model of \cite{babu1} based on
$SU(5)\times A_4$ including the
SSB sector is investigated in Sect. III,
where $A_4$ is the group of even permutations
of four objects.
As we will see, the model has a
strong predictive power, and we
will calculate various low-energy parameters
such as the top quark mass and the spectrum of
the superpartners that
are predicted from the model.
In Sect. IV
we conclude.

\section{Softly broken $N=1$ supersymmetric  Finite Unified Theories}

We start by considering a generic form of
the superpotential \footnote{
We follow the notation of \cite{jack4}.}
\be
W &=&\frac{1}{6} \,Y^{ijk}\,\Phi_i \Phi_j \Phi_k
+\frac{1}{2} \,\mu^{ij}\,\Phi_i \Phi_j,
\ee
along with the Lagrangian for the SSB terms
\be
-{\cal L}_{\rm SB} &=&
\frac{1}{6} \,h^{ijk}\,\phi_i \phi_j \phi_k
+
\frac{1}{2} \,B^{ij}\,\phi_i \phi_j
+
\frac{1}{2} \,(m^2)^{j}_{i}\,\phi^{*\,i} \phi_j+
\frac{1}{2} \,M\,\lambda \lambda+\mbox{H.c.},
\ee
where $\phi_i$ is the scalar component of $\Phi_i$, and
$\lambda$ stands for gaugino.
Since we  consider
only finite theories, we assume that
the one-loop
$\beta$ function of the
gauge coupling $g$ vanishes, i.e.,
\be
\sum_i T(R_i)-3 C(G)=0,
\label{Q}
\ee
where $T(R_i)$ is the Dynkin index of the representation $R_i$ and $C(G)$
 is the
quadratic Casimir of the adjoint representation of the
gauge group $G$. 
We also assume that the gauge group is  a simple group,
and that the theory is free from the gauge anomaly, of course.
According to the finiteness theorem of \cite{lucchesi1},
the theory is  finite
to all orders in perturbation theory \footnote{Finiteness here
means only for dimensionless couplings
$g$ and $Y^{ijk}$.}, if \\
(i) the reduction equation \cite{zim1,kubo,oehme1}
\be
\beta_{Y}^{ijk} &=& \beta_{g}\,d Y^{ijk}/d g
\label{Yg2}
\ee
admits a unique power series solution
\be 
Y^{ijk} &=& g\,\sum_{n=0}\,\rho^{ijk}_{(n)} g^{2n},
\label{Yg}
\ee
where $\beta_{g}$ and $ \beta_{Y}^{ijk}$
are the $\beta$ functions of $g$ and $Y^{ijk}$, respectively
\footnote{See \cite{kubo0} for further  references on
reduction of couplings.}, and\\
(ii)
the one-loop anomalous dimensions
 vanish, that is, 
\be
 \frac{1}{2}\sum_{p,q}\rho_{ipq(0)} \rho^{jpq}_{(0)}
-2\delta^{j}_{i}\,C(R_i) &=&0,
\label{P}
\ee
 where $\rho_{ipq(0)} = \rho^{* jpq}_{(0)}$.
We would like to recall that if the condition (ii)
is satisfied, the two-loop expansion coefficients
in (\ref{Yg}), $\rho^{ijk}_{(1)}$, vanish \cite{jack3}, and that
if (i) and (ii) are satisfied, the anomalous dimensions
$\gamma_{i}^{j}$ vanish to all orders \cite{lucchesi1}.
Field theories that satisfy (i) and (ii) possess the exact
scale invariance.

In the presence of the SSB terms,
the exact
scale invariance is broken by them
in a strict sense.
However, it is expected that the couplings, masses etc in
a unified field theory without gravity  are VEV's of certain
fields in a more fundamental theory.
Therefore, it would be natural to transform them
under the  scale transformation, too.
Then the scale invariance of a 1PI function $\Gamma$
means:
\be
\Gamma[e^t p, e^{t}h, e^{t}\mu, e^{2 t} B, e^{2t}m^2,Y,g] &=&
e^{d_\Gamma t}\Gamma[p, h, \mu,  B,
m^2,Y,g] ,
\label{scale}
\ee
where $p$ stand for momenta, and $d_\Gamma$ is the canonical dimension
of $\Gamma$. Clearly, (\ref{scale}) is correct, only if the theory is
finite.
Finiteness in the SSB sector can be achieved
by using the relations among the  renormalization of
the SSB parameters and those of an unbroken
supersymmetric gauge theory
\cite{kazakov3,kubo4,jack4,yamada,hisano,jack5,kazakov4,giudice1,jack6,kazakov5,kraus}
\footnote{In \cite{terao} this matter is
reviewed in a transparent way. See also \cite{ns1}.}.
Accordingly, the $\beta$ functions of the $M, h$ and $m^2$
parameters  can be written as \cite{jack5,kazakov4}
\footnote{We do not consider $B^{ij}$
in the following discussions, because they do not enter into the
$\beta$ functions of the other quantities \cite{jack5,kazakov4}.
Moreover, they are
automatically finite if the other quantities are finite \cite{kazakov3}.}
\be
\beta_M &=& 2{\cal O}\left({\beta_g\over g}\right),
\label{betaM}\\
\beta_h^{ijk}&=&\gamma^i{}_lh^{ljk}+\gamma^j{}_lh^{ilk}
+\gamma^k{}_lh^{ijl}-2\gamma_1^i{}_lY^{ljk}
-2\gamma_1^j{}_lY^{ilk}-2\gamma_1^k{}_lY^{ijl},\\
(\beta_{m^2})^i{}_j &=&\left[ \Delta
+ X g \frac{\partial}{\partial g}\right]\gamma^i{}_j,
\label{betam2}\\
{\cal O} &=&\left(Mg^2{\partial\over{\partial g^2}}
-h^{lmn}{\partial
\over{\partial Y^{lmn}}}\right),
\label{diffo}\\
\Delta &=& 2{\cal O}{\cal O}^* +2|M|^2 g^2{\partial
\over{\partial g^2}} +\tilde{Y}_{lmn}
{\partial\over{\partial
Y_{lmn}}} +\tilde{Y}^{lmn}{\partial\over{\partial Y^{lmn}}},
\label{delta}\ee
where $(\gamma_1)^i{}_j={\cal O}\gamma^i{}_j$,
$Y_{lmn} = (Y^{lmn})^*$, and
\be
\tilde{Y}^{ijk}&=&
(m^2)^i{}_lY^{ljk}+(m^2)^j{}_lY^{ilk}+(m^2)^k{}_lY^{ijl}.
\ee
$X$ in (\ref{betam2}) has been first explicitly calculated
in the lowest order \cite{jack3,jack7}, and later its all order form
\cite{kubo4,jack6,kazakov5,kraus,terao}
\be
X &=&
\frac{-|M|^2 C(G)+\sum_\l m_\l^2 T(R_\l) }{C(G)-8\pi^2/g^2}~
\label{xtilde2}
\ee
has been found
in the renormalization scheme of Novikov {\em et al.},
in which the $\beta$ function of the gauge coupling $g$ is given by
\cite{novikov1} 
 \be
\beta_g^{\rm NSVZ} &=&
\frac{g^3}{16\pi^2}
\left[ \frac{\sum_\l T(R_\l)(1-2\gamma_\l)
-3 C(G)}{ 1-g^2C(G)/8\pi^2}\right].
\label{bnsvz}
\ee
The key point in 
\cite{kazakov3,kubo4,jack4}
is the assumption that the differential operators
${\cal O}$ and $\Delta$ given in (\ref{diffo})
and  (\ref{delta})
become total derivative operators
on the RG invariant surface
which is defined by the solution
of the reduction equations for the SBB parameters.
It has been shown in \cite{kazakov3,kubo4,jack4} that if the trilinear
couplings
are expressed in terms of $M$ and $g$ as \cite{kazakov3,jack4}
\footnote{Reduction of massive parameters has been first proposed in
\cite{kubo1}.}
\be
h^{ijk} &=&-M \frac{d Y^{ijk}(g)}{d \ln g},
\label{h}
\ee
and the soft scalar masses satisfy the sum rules \cite{kubo4}
\be
m^2_i+m^2_j+m^2_k &=&
|M|^2 \{~
\frac{1}{1-g^2 C(G)/(8\pi^2)}\frac{d \ln Y^{ijk}}{d \ln g}
+\frac{1}{2}\frac{d^2 \ln Y^{ijk}}{d (\ln g)^2}~\}\nn\\
& +&\sum_\l
\frac{m^2_\l T(R_\l)}{C(G)-8\pi^2/g^2}
\frac{d \ln Y^{ijk}}{d \ln g},
\label{sum2}
\ee
the differential operators ${\cal O}$ and $\Delta$
become total derivative with respect to $g$:
\be
{\cal O} &=& \frac{M}{2}
\frac{d }{d \ln g},
\label{diffo2}\\
\Delta +X g\frac{\partial}{\partial g}&=&
|M|^2 \{~
\frac{1}{1-g^2 C(G)/(8\pi^2)}\frac{d }{d \ln g}
+\frac{1}{2}\frac{d^2 }{d (\ln g)^2}~\}\nn\\
& +&\sum_\l
\frac{m^2_\l T(R_\l)}{C(G)-8\pi^2/g^2}
\frac{d }{d \ln g}.
\label{delta2}
\ee
Note that in the derivations from (\ref{h}) to (\ref{delta2}), it has been
assumed
that
\be
\gamma^j{}_i &=& \gamma_i \delta^j{}_i,
\label{as1}\\
(m^2)^j{}_i &=& m^2_i \delta^j{}_i,
\label{as2}\\
Y^{ijk}\frac{\partial}{\partial Y^{ijk}}
&=& Y^{*ijk}\frac{\partial}{\partial Y^{*ijk}}~
\mbox{on the space of the RG functions}.
\label{as3}
\ee
Therefore, if the anomalous dimensions $\gamma_i $
vanish to all orders ( which is ensured if (i) and (ii) given in
(\ref{Yg}) and 
(\ref{P}) are satisfied), we have:
$\beta_M=\beta_h^{ijk}=(\beta_{m^2})^{i}_{j}=0$.

We see from (\ref{sum2}) that
the universal choice
\be
m_{i}^{2} &=&\frac{|M|^2}{3}
\label{sym}
\ee
also ensure the finiteness to two-loop order
in accord with\cite{jones1,jack3,kazakov3}. Note
that $C(G)=\sum_\l T(R_\l) /3$ and
$d \ln Y^{ijk}/d \ln g =1 +0(g^4)$. Similarly,
the $N=4$ supersymmetric case ($T(R_\l)=C(G)$)
with the SSB parameters \cite{parkes1}
can be simply derived from (\ref{h}) and (\ref{sum2}).

To summarize, finiteness in the SSB sector is guaranteed if
$h^{ijk}$ are expressed according to (\ref{h}), and the
sum rules (\ref{sum2}) are satisfied.
The trilinear couplings $h^{ijk}$, unless they are aligned, contribute
to $\delta_{RL}$ \cite{fcnc} which
are strongly constrained from
FCNC processes and dangerous CP-violating phenomena.
The explicit form of $h^{ijk}$ in finite theories is known to
two-loop order \cite{jones1,jack3}:
\be
h^{ijk}=-M Y^{ijk}(g)+O(g^5).
\ee
The higher order terms depend on the renormalization scheme.
In fact,  it is possible \cite{oehme1} to
make vanish  all the expansion coefficients
$\rho_{(n)}^{ijk}$ of (\ref{Yg}) except
the lowest order one $\rho_{(0)}^{ijk}$
by a suitable redefinition of the Yukawa couplings $Y^{ijk}$.
The redefinition does not modify
the  form of $\beta$ function $\beta_g^{\rm NSVZ} $
(\ref{bnsvz}), because only the anomalous dimensions
change in $\beta_g^{\rm NSVZ}$.
Therefore, in finite theories, $h^{ijk}$ are aligned to
all orders, and therefore,
$h^{ijk}$ introduce no extra CP-violating phases:
\be
h^{ijk}=-M Y^{ijk}(g).
\label{hijk}
\ee
In contrast to this case, the sum rules (\ref{sum2})
of the soft scalar masses do not automatically
ensure  their  degeneracy
in the space of generation.
However, as we will see in a concrete model,
the sum rules (\ref{sum2}) can yield
the  degeneracy of the soft scalar masses.
The exact finiteness does not automatically
yield a solution to the SUSY flavor problem.
But a solution to the SUSY flavor problem
can result from the quantum scale invariance.

Since superstring theories are scale invariant theories,
a solution to  the SUSY flavor problem
based on the exact scale invariance
may be realized.
In fact, in a certain class of orbifold models of
superstrings,
in which the massive string states are organized
into $N = 4$ supermultiplets \cite{DKL2} (see also \cite{IL}),
so that they do not contribute to
the quantum modification of the kinetic
function, 
the sum rules
\be
m^2_i+m^2_j+m^2_k &=&
|M|^2~
\frac{1}{1-g^2 C(G)/(8\pi^2)}+\sum_\l
\frac{m^2_\l T(R_\l)}{C(G)-8\pi^2/g^2}~
\label{sum31}
\ee
along with $h^{ijk}=-M Y^{ijk}$ are satisfied  \cite{kubo5}.
(See also \cite{BIM1,kobayashi,BIM2,SD,kkk,kim1}.)
Therefore,
the finiteness conditions (\ref{hijk}) and (\ref{sum2})
coincide with those of
the above superstrings models to all orders.

\section{Model based on $SU(5)\times A_4$}

There exist various unified models that are all-order finite
at least in the
dimensionless sector \cite{lucchesi1,kapetanakis,kazakov2,yoshioka1,kubo5}.
In all the models, only such solutions of the
reduction equations(\ref{Yg2}) have been considered that
admit
the hierarchal structure of the Yukawa couplings.
As a result, the sum rules (\ref{sum2})
are indeed satisfied, but
it cannot strongly
constrain the individual soft scalar masses
in the first two generations (See \cite{kubo5}).
In contrast to the previous models, the
$SU(5)$ models (two of three models) proposed
in \cite{babu1} yield
a democratic structure
of the Yukawa couplings.
As we will see, the democratic structure
(which follows as a consequence of
certain discrete symmetries) is essential to obtain
sum rules of the soft scalar masses from which
their degeneracy  follows.

Three
generations of quarks and leptons   are accommodated in
${\bf 10}_i$ and $\overline{\bf 5}_i$,
where $i$ runs over the three generations.
A $\Sigma$ in {\bf 24} is used to break $SU(5)$ down to $SU(3)_{\rm C}
\times SU(2)_{\rm L} \times U(1)_{\rm Y}$,  and
there are four pairs of Higgs
supermultiplets $H_a$ and $\overline{H}_a~~(a=1 \sim 4)$.
The starting superpotential
 is 
\be
W &=& 
\sum_{i,j=1}^{3}\sum_{a=1}^4\left( \frac{1}{2}u_{ij}^a\,
{\bf 10}_i {\bf 10}_j H_a+
d^a_{ij}\, {\bf 10}_i \overline{\bf 5}_j \overline{H}_a \right)\nn\\
&+ &
\sum_{a,b=1}^4\kappa^{a b}\, \overline{H}_a
\Sigma H_b+\frac{\lambda}{3}\Sigma^3
+ \frac{\mu_{\Sigma}}{2}\,
\Sigma^2+ \mu_{H}^{a b}\,\overline{H}_a H_{b},
\label{superp}
\ee
and the SSB Lagrangian
is
\be
-{\cal L}_{\rm SSB} &=&
\sum_{a=1}^4 [~m_{H_a}^{2}\hat{H}^{*}_a \hat{H}_{a}
+m_{\bar{H}_a}^{2}
\hat{\overline {H}}^{*}_{a}\hat{\overline {H}}_{a}~]
+m_{\Sigma}^{2}{\hat\Sigma}^{\dag}
{\hat\Sigma} \nn\\
& +&
\sum_{i=1}^{3}[~m^2_{5_i}~\hat{\overline{\bf 5}}^{*}_i
\hat{\overline{\bf 5}}_i+
m^2_{10_i}\hat{{\bf 10}}_i^{*}\hat{{\bf 10}}_i~]\nn\\
& +&
\left\{ \,
 \frac{1}{2}M\lambda \lambda+B_{\Sigma}{\hat \Sigma}^{2}+
\sum_{a,b=1}^4[~B_H^{ab}\hat{\overline {H}}_{a}{\hat H}_{b}
+h_{f}^{ab}\,\hat{\overline{H}}_{a}
{\hat \Sigma} {\hat H}_{b}~] \right.\nn\\
& +&\left. \frac{h_{\lambda}}{3}\,{\hat \Sigma}^3+
\sum_{i,j=1}^{3}\sum_{a=1}^4\left( \frac{h_{u\, ij}^a}{2}\,
\hat{{\bf 10}}_i \hat{{\bf 10}}_j \hat{H}_a+
h^a_{d\, ij}\, \hat{{\bf 10}}_i \hat{\overline{\bf 5}}_j
\hat{\overline{H}}_a\right)
+\mbox{h.c.}\, \right\},
\label{Lssb}
\ee
where a hat is used to denote the scalar
component of each chiral supermultiplet.
The resulting theory has an unbroken R-parity along with the
conservation of $B-L$.
Note that we assumed the diagonal soft scalar masses, because
non-diagonal soft masses would not satisfy
the assumption (\ref{P}) as well as (\ref{as1}),
and hence violates finiteness.

\subsection{The degeneracy of the soft scalar masses from
their sum rules}
To be specific, we consider
the model  based on $SU(5)\times A_4$ symmetry \cite{babu1},
where $A_4$ is the group of even permutations
\footnote{The $S_4$ model \cite{babu1} can be treated similarly.
We have found that as far as the SSB sector is concerned,
it is exactly the same as the $A_4$ model.}.
$A_4$ has three irreducible representations
${\bf 1}~,~{\bf 1}'~,~{\bf 1}''$ and ${\bf 3}$  \cite{ma}, and
the matter supermultiplets belong to its
representation according to \cite{babu1}
\begin{eqnarray}
{\bf 10}_i &:& {\bf 3}~,~\overline{{\bf 5}}_i ~:~
{\bf 3}\nonumber\\
(H_i,H_4)  &:& {\bf 3}+{\bf 1}'~,~
(\overline{H}_i, \overline{H}_4)  ~:~ {\bf 3}+{\bf 1}''~~~~\Sigma
~:~{\bf 1},\nonumber
\end{eqnarray}
where $i=1 \sim 3$. Then the cubic part of the superpotential (\ref{superp})
invariant under $A_4$ becomes
\begin{eqnarray}
W_3&=&\frac{a}{2}({\bf 10}_1 {\bf 10}_1
+\omega {\bf 10}_2 {\bf 10}_2 +\omega^2 {\bf 10}_3
{\bf 10}_3)H_4\nonumber\\
&+&c({\bf 10}_1 \overline{{\bf 5}}_1
+\omega^2 {\bf 10}_2 \overline{{\bf 5}}_2
+\omega  {\bf 10}_3\overline{{\bf 5}}_3)\overline{H}_4
\nonumber\\
&+&b{\bf 10}_1 {\bf 10}_2H_3
+d({\bf 10}_1 \overline{{\bf 5}}_2 +{\bf 10}_2
\overline{{\bf 5}}_1)\overline{H}_3
\nonumber\\
&+&b{\bf 10}_3 {\bf 10}_1H_2
+d({\bf 10}_3 \overline{{\bf 5}}_1 +{\bf 10}_1 \overline{{\bf 5}}_3 )
\overline{H}_2\label{sp2}\\
&+&b{\bf 10}_2 {\bf 10}_3H_1+d({\bf 10}_2 \overline{{\bf 5}}_3
+{\bf 10}_3 \overline{{\bf 5}}_2)\overline{H}_1\nonumber\\
&+&k(\overline{H}_{1}H_{1}+\overline{H}_{2}H_{2}
+\overline{H}_{3}H_{3})\Sigma
+\frac{\lambda}{3}\Sigma^3,\nonumber
\end{eqnarray}
where $w=\exp(i2\pi/3)$ can be removed by field redefinition.
The lowest order solution to the reduction equation (\ref{Yg2}) is
\cite{babu1}:
\be
a^2=b^2=\frac{8}{15}g^2~,~
c^2=d^2=e^2=\frac{2}{5}g^2~,~k^2=\frac{1}{3}g^2~,~\lambda^2=
\frac{15}{7} g^2.
\label{solutionA}
\ee
It can be shown that
the power series solution (\ref{Yg}) exists uniquely, so that the
dimensionless
sector can be made finite to any finite order in perturbation theory.
At this point we assume that a suitable redefinitions of the
Yukawa couplings \cite{oehme1} has been performed so that the above solution
(\ref{solutionA}) is exact.
Note  the mass term for the $H$ and $\overline{H}$
is not invariant under $A_4$ for an arbitrary mass matrix $\mu_H$.
The choice of $\mu_H$ is however very important to make
the model phenomenologically viable, because
the Cabibbo-Kobayashi-Maskawa (CKM) mixing
of the quarks in this model basically
originates from  $\mu_H$.
So, the $A_4$ invariance has to be
broken by the mass term, already at the GUT scale $M_{\rm GUT}$.
Therefore, it is natural to assume that the
operators with  dimension less than four
do not have to respect the $A_4$ invariance.
Since the SSB terms consist of
such operators, we should not
impose the  $A_4$ invariance on the SSB Lagrangian  (\ref{Lssb}).
We proceed with this remark in mind.

Eq. (\ref{hijk})  means
\be
h_{u~ij}^a &=& -M u_{ij}^a~,~
h_{d~ij}^a = -M d_{ij}^a.
\label{hA}
\ee
Further, the right hand side of (\ref{sum2})
(which we denote by $\tilde{M}^2$) can be written as
\be
\tilde{M}^2 &=&
|M|^2 
\frac{1}{1-g^2 C(G)/(8\pi^2)}+\sum_\l
\frac{m^2_\l T(R_\l)}{C(G)-8\pi^2/g^2}.
\label{Mtilde}
\ee
Using $\tilde{M}^2$ above,  we write down all the sum rules
(\ref{sum2}):
\be
\tilde{M}^2 &=& 2 m^2_{10_1}+m^2_{H_4}
=2 m^2_{10_2}+m^2_{H_4}=2 m^2_{10_3}+m^2_{H_4},
\label{sum41}\\
\tilde{M}^2 &=& m^2_{10_1}+m^2_{10_2}+m^2_{H_1}=
m^2_{10_1}+m^2_{10_3}+m^2_{H_2}=
m^2_{10_2}+m^2_{10_3}+m^2_{H_3} ,\label{sum42}\\
\tilde{M}^2 &=& m^2_{10_1}+m^2_{\bar 5_1}+m^2_{\bar H_4}=
 m^2_{10_2}+m^2_{\bar 5_2}+m^2_{\bar H_4}=
m^2_{10_3}+m^2_{\bar 5_3}+m^2_{\bar H_4},
\label{sum48}\\
\tilde{M}^2 &=& m^2_{10_1}+m^2_{\bar 5_2}+m^2_{\bar H_1}=
m^2_{10_2}+m^2_{\bar 5_1}+m^2_{\bar H_1},
\label{sum43}\\
\tilde{M}^2 &=& m^2_{10_1}+m^2_{\bar 5_3}+m^2_{\bar H_2}=
m^2_{10_3}+m^2_{\bar 5_1}+m^2_{\bar H_2},
\label{sum44}\\
\tilde{M}^2 &=& m^2_{10_2}+m^2_{\bar 5_3}+m^2_{\bar H_3}=
m^2_{10_3}+m^2_{\bar 5_2}+m^2_{\bar H_3},
\label{sum45}\\
\tilde{M}^2 &=& m^2_{H_1}+m^2_{\bar H_1}+m^2_{\Sigma}=
m^2_{H_2}+m^2_{\bar H_2}+m^2_{\Sigma}=
m^2_{H_3}+m^2_{\bar H_3}+m^2_{\Sigma},
\label{sum46}\\
\tilde{M}^2 &=& 3 m^2_{\Sigma}.\label{sum47}
\ee
The sum rules (\ref{sum41}) require the degeneracy of $ m^2_{10_i} $,
and the degeneracy of
$m^2_{\bar 5_i}$ follows from (\ref{sum48}).
Similarly, one can easily derive
the degeneracy of 
$m^2_{\bar{H}_a}$ as well as that of $m^2_{H_a}$:
\be
m^2_{10_i} &=& m^2_{10}~,~
m^2_{\bar 5_i} =  {4 \over 3 } \tilde{M}^2  - 3 m^2_{10} ,\nn\\
m^2_{H_a} & = &  \tilde{M}^2 - 2  m^2_{10}~,~
m^2_{\bar H_a} = - {1 \over 3} \tilde{M}^2 + 2  m^2_{10} ,
\label{sum3}
\ee
where $i=1 \sim 3$ and $a =1\sim 4$. As we can see from (\ref{sum3}),
There are only two independent parameters in the SSB sector,
 $m^2_{10}$ and the gaugino mass M for instance
 as we have indicated in (\ref{sum3}).
To express $\tilde{M}^2$ in terms of $M$, we have to compute the trace
$  $ in (\ref{Mtilde}). We find:
\be
m^2_\l T(R_\l) &=&
\frac{1}{2}\left[\sum_{i=1}^3 m^2_{\bar 5_i}+
\sum_{a=1}^4 (m^2_{H_a}+m^2_{\bar H_a})\right]+
\frac{3}{2}\sum_{i=1}^3 m^2_{10_i}+5 m^2_{\Sigma}\nn\\
&=& C(SU(5))\tilde{M}^2,
\ee
where we have used (\ref{sum41})--(\ref{sum47}).
Using this, we then obtain
\be
\tilde{M}^2 &=& |M|^2.
\ee

Note that the democratic structure for the quark
mass matrices is essential to obtain the set of
 the sum rules (\ref{sum48}) and
 (\ref{sum43})--(\ref{sum45}) that
yields the universal soft masses (\ref{sum3}).
To summarize, finiteness requires that
the trilinear couplings have to be aligned (\ref{hA}) and the soft scalar
masses
have to have the universal form (\ref{sum3}).
Before the diagonalization of $\mu_H^{ab}$,
the Yukawa couplings $u_{ij}$ and $d_{ij}$ are real numbers. Note that
there are no restrictions on $\mu_H^{ab}$ and $B_H^{ab}$ from finiteness.
The diagonalization of $\mu_H^{ab}$ and  an appropriate phase rotation
of $H_a$ and $\overline{H}_a$ will introduce
phases into the Yukawa couplings, which yields
the ordinary CKM phase.
The redefinition of the superfields above does not
destroy the alignment of the trilinear couplings (\ref{hA}) and
the universality of the  soft masses (\ref{sum3}).
Then only the gaugino mass $M$ and $B_H^{ab}$ are
complex numbers and contain CP-violating phases
in this model. They may contribute to the EDM
of the neutron, for instance \cite{fcnc-edm}.
Nevertheless, the SUSY flavor problem is
drastically reduced in this finite unified model.

\subsection{Predictions at low-energy }

Since there are four pairs of the Higgs supermultiplets,
it is not all automatic that there is only one pair of light Higgs
doublets at low energies after a fine tuning at $M_{\rm GUT}$.
Furthermore, the Yukawa couplings are of oder $O(g)$,
and so we have to worry about the problem of fast proton decay
\cite{deshpande} 
via dimension five operators \cite{d5}.
These problems are related to the choice of the
supersymmetric Higgs mass matrix $\mu_H^{ab}$ in
the superpotential (\ref{sp2}), which we would
like to  leave for feature problem. Note that
there are no constraints on $\mu_H^{ab}$
from finiteness.
In what  follows, we simply assume that
there is one pair of  light Higgs
doublets and the proton decay can be sufficient suppressed.

The finiteness conditions (\ref{solutionA}),   (\ref{hA}) and
(\ref{sum3})
do not restrict the renormalization property at low energies,
because the gauge symmetry is spontaneously broken
below $M_{\rm GUT}$. This should be contrasted to the case
of the anomaly mediated supersymmetry breaking \cite{anomaly}.
Therefore, the conditions (\ref{solutionA}),
(\ref{hA}) and (\ref{sum3}) are just boundary conditions at $M_{\rm GUT}$
in our case.
By assumption,  the evolution of the parameters below $M_{\rm GUT}$ is
governed by the MSSM \cite{rgf}. We further assume a unique
supersymmetry breaking scale
$M_{\rm SUSY}$,
which is identified  with $\sqrt{(m^2_{t_1}+m^2_{t_2})/2}$,
where $m^2_{t_i}$ are two
stop masses,  so that
below $M_{\rm SUSY}$ the SM is the correct effective theory.
We  recall that
$\tan\beta$ is usually determined in the Higgs sector.
However, 
in the case at hand, it is convenient
to define $\tan\beta$ by using
the matching condition at $M_{\rm SUSY}$,
\be
\alpha_{t}^{\rm SM}
&=&\alpha_{t}\,\sin^2 \beta~,~
\alpha_{b}^{\rm SM}
~ =~ \alpha_{b}\,\cos^2 \beta~,~
~\alpha_{\tau}^{\rm SM}
~=~\alpha_{\tau}\,\cos^2 \beta,\nn\\
\alpha_{\lambda}&=&
\frac{1}{4}(\frac{3}{5}\alpha_{1}
+\alpha_2)\,\cos^2 2\beta,
\label{match}
\ee
where $\alpha_{i}^{\rm SM}~(i=t,b,\tau)$ are
the SM Yukawa couplings and $\alpha_{\lambda}$ is
the Higgs coupling $(\alpha_I=g^{2}_{I}/4\pi^2)$.
The matching conditions (\ref{match}) and the
boundary conditions at $M_{\rm GUT}$ can be satisfied only for a specific
value of $\tan\beta$.
This is the reason of why it is possible
without knowing the
details of the scalar sector of the MSSM
to predict various 
parameters such as
 the top and quark 
masses \cite{kapetanakis,yoshioka1,kubo2}.

Since $\tan\beta$ is fixed in the dimension-zero sector and
the soft scalar masse have to satisfy the boundary conditions
(\ref{sum3}) at $M_{\rm GUT}$,
it is by no means trivial that
the electroweak symmetry is correctly broken
at low energies \cite{inoue1}.
Fortunately,  the supersymmetric mass parameter
$\mu_H$ for the pair of the light Higgs doublets and the corresponding
$B$ term 
are not constrained by the finiteness conditions.
Therefore, we use this
freedom to fix $\mu_H$ and $B$
to trigger the electroweak symmetry
breaking. 
To proceed we write down
the up-quark mass matrix at $M_{\rm GUT}$
which can be read off from (\ref{sp2}) and (\ref{solutionA}):
\be
M^u &=&
\sqrt{\frac{8}{15}}~g <H_4>~\left(
\begin{array}{ccc} 1 & 1+\epsilon_1 &1+\epsilon_2\\
1+\epsilon_1&1 &1+\epsilon_3 \\
1+\epsilon_2 &1+\epsilon_3 & 1
 \end{array}\right),
 \label{matrix}
\ee
where
\be
\epsilon_i &=& \frac{<H_i>}{<H_4>}-1~~\mbox{with}~~~i=1,2,3.
\ee
As we can see also from (\ref{sp2}) and (\ref{solutionA}),
the down-type quark mass matrix
has the same structure.
It has been found \cite{fishbane} that
the above democratic mass matrix with
 $\epsilon_i << 1$ agree with experimental data.
 Therefore, $H_a$ have to have almost equal
VEV's, although there is only one
pair of light Higgs doublets.

After so much remarks, we are now in position to
compute low energy quantities.
We use the renormalization group equations of two-loop order
for dimensionless parameters and
 those of one-loop order
for dimensional ones \cite{rgf}. To see the gross nature
of the low energy predictions of the present model, we however neglect
$\epsilon_i$ in the mass matrix (\ref{matrix}), and the
the threshold corrections at $M_{\rm GUT}$ as well as $M_{\rm SUSY}$,
while we take into account the SM correction to
the physical mass of the top quark $m_t$, and $m_b$
which is the running bottom mass at $m_b$.
Under this simplification,
the top and bottom Yukawa couplings
at $M_{\rm GUT}$ are given by
\be
y_t &=& \sqrt{6/5}~g~,~y_b=\sqrt{9/10}~g.
\ee
In Table I, we present the predictions
for $\alpha_{3}(M_Z), \tan\beta, m_t$ and $m_b$ for different choices
of the unified gaugino mass $M$.
Comparing, for instance,
the $m_t$ prediction above
 with the most recent
experimental value  \cite{topmass},
\be
m_t = (174.3 \pm 5.1) ~~\mbox{GeV },
\ee
and recalling that we have neglected
$\epsilon_i$ in (\ref{matrix}), and the
the threshold corrections,
we see that the prediction can be  consistent with the experimental
data.

Next we turn to the SSB sector with the
finiteness conditions (\ref{hA}) and (\ref{sum3}).
As we can see from (\ref{sum3}),
we may treat $M$ and $m_{10}$ as independent parameters.
The nice feature of (\ref{sum3}) is
that the soft scalar masses
of $H_a$ and $\overline{H}_a$ are  degenerate.
Therefore,  one pair  of the light Higgs doublets,
$H_u$ and $H_d$, 
which can be obtained after the diagonalization
by an appropriate  unitary matrix, has exactly the
same soft scalar mass.
Then we look for the parameter space in the  $m_{10}-M$ plane,
in which a successful
radiative electroweak symmetry breaking occurs
and the lightest neutralino is the LSP.
In Fig. 1,  we show the result,
where the region with dots and open  squares  leads to a successful
radiative electroweak symmetry breaking. In the region with
dots, the lightest  neutralino is the LSP, while in
the region with open  squares the LSP is the stau.
So the phenomenologically viable   parameter space
in the $M-m_{10}$ plane is very restricted;
$m_{10}$ has to lie approximately on the straight line
given by $m_{10}=5/8 M$.
That is, for a given unified gaugino mass $M$, the spectrum
of the superpartners is basically fixed.
In Table II we present the results for
$M=1$ TeV and $1.5$ TeV in an obvious notation.

The dotted region in Fig. 1 is interesting also from the 
cosmological viewpoint.
In the dotted region, the lightest neutralino $\chi_1$ 
and the light stau $\tilde \tau_1$ are nearly degenerate 
in mass, that is, $m_{\tilde \tau_1} - m_{\chi_1} < 25$ GeV, 
where the light stau $\tilde \tau_1$ is the next-to-LSP.
With this type of spectrum, neutralino-stau co-annihilation can 
occur and that reduces the relic LSP density \cite{dark-matter}.
Thus, this parameter region is quite interesting 
for the LSP dark matter scenario.

\section{Conclusion}

Although it was suggested in past
\cite{parkes1,jones1,jack3,kazakov3} that
finiteness of softly broken supersymmetric
Yang-Mills theories
may play an important role to
understand the universality of the SSB parameters,
there was so far no finite model based on softly broken $N=1$
supersymmetry
in which the universality of the SSB parameters
follows solely from finiteness.
The simple reason is the relaxed finiteness condition,
the sum rule (\ref{sum2}).

In this paper we considered the finite model of
\cite{babu1} which is based on the discrete symmetry
$A_4$ and has the democratic structure of the
mass matrices for the quarks and leptons.
We included the SSB sector to this model
and required that this sector does not destroy
finiteness. $A_4$ symmetry in the SSB sector was
not assumed, because $A_4$ symmetry has to be broken
at $M_{\rm GUT}$ by operators
with dimension less than four.
We found that
finiteness in this model requires that
the trilinear couplings have to be aligned (\ref{hA}) and the soft scalar
masse
have to have the universal form (\ref{sum3}).
The democratic structure of the
mass matrices played the essential role to obtain
the universality of the soft scaler masses.
Therefore, this model shows that finiteness can
offers a solution to the SUSY flavor problem, and
indicates that
the SUSY flavor problem is closely related to
the exact scale invariance.

\acknowledgments

We would like to than M. Mondragon for useful discussions.
This work is supported by the Grants-in-Aid
for Scientific Research  from
 the Japan Society for the Promotion of Science (JSPS)
 (No. 14540252,13135210,  14540256).
 The work of KB is supported in part by DOE Grant \# DE-FG03-98ER-41076,
a grant from the Research Corporation and by DOE Grant
\# DE-FG02-01ER-45684.
 This work was partially conducted by
 way of a grant awarded by the Government of Mexico in the
 Secretariat of Foreign Affairs.

\begin{table}
\caption{The predictions
for different $M$.}
\begin{center}
\begin{tabular}{|c|c|c|c|c|}
$M$ [TeV]   &$\alpha_{3(5{\rm f})}(M_Z)$ &
$\tan \beta$ 
 & $m_{b}$ [GeV]& $m_{t}$ [GeV]
\\ \hline
$1$ & $0.118 $  &$52 $  & $4.6$
 & 179\\ \hline
$1.5$ & $0.117 $  &$52 $  & $4.6$
  & 179 \\ 
\end{tabular}
\end{center}
\end{table}

\begin{table}
\caption{The predictions
of the spectrum of the superpartners
for $M=1$ TeV with $m_{10}=  0.63 $ TeV and $M=1.5$ TeV
with $m_{10}=  0.94 $ TeV.}
\begin{center}
\begin{tabular}{|c|c||c|c|}
 \hline
$m_{\chi_1}$ (TeV) &
0.45~~~~0.69 &
$m_{\tilde{s}_1}=m_{\tilde{d}_1}$ (TeV)  &
1.9~~~~2.8 \\ \hline
$m_{\chi_2}$ (TeV) &
0.84~~~~1.3 &
$m_{\tilde{s}_2}=m_{\tilde{d}_2}$ (TeV)  &
2.2~~~~3.2
\\ \hline
$m_{\chi_3} $ (TeV) &
1.3~~~~1.9 &
$m_{\tilde{\tau}_1}$ (TeV)   &
0.43~~~~0.69
\\ \hline
$m_{\chi_4}$ (TeV) &
1.3~~~~1.9 &
$m_{\tilde{\tau}_2}$ (TeV)   &
0.72~~~~1.0
\\ \hline
$m_{\chi^{\pm}_{1}} $ (TeV) &
0.84~~~~1.3 &
$m_{\tilde{\mu}_1}=m_{\tilde{e}_1}$ (TeV) &
0.78~~~~1.2
\\ \hline
$m_{\chi^{\pm}_{2}} $ (TeV) &
1.3~~~~1.9 &
$ m_{\tilde{\mu}_2}=m_{\tilde{e}_2}$
(TeV) & 
1.1~~~~1.6
\\ \hline
$m_{\tilde{t}_1}$ (TeV) &
1.5~~~~2.2 &
$ m_{\tilde{\nu_\tau}}$
(TeV) &
0.68~~~~1.0
\\ \hline
$m_{\tilde{t}_2}$  (TeV) &
1.7~~~~2.5 &
$m_{\tilde{\nu_\mu}}=m_{\tilde{\nu_e}}$
(TeV) & 0.78~~~~1.2
\\ \hline
$m_{\tilde{b}_1}$  (TeV) &
1.5~~~~2.2  & 
$m_{A}$  (TeV) &
0.62~~~~0.93 
\\  \hline
$m_{\tilde{b}_2}$  (TeV) &
1.7~~~~2.5  & 
$m_{H^\pm}$  (TeV) &
0.63~~~~0.94 
\\  \hline
$m_{\tilde{c}_1}=m_{\tilde{u}_1}$  (TeV) &
2.1~~~~3.1  & 
$m_{H}$  (TeV) &
0.63~~~~0.93  
\\  \hline
$m_{\tilde{c}_2}=m_{\tilde{u}_2}$  (TeV) &
2.2~~~~3.2  & 
$m_{h}$  (TeV) &
0.13~~~~0.13 
\\  \hline
\end{tabular}
\end{center}
\end{table}

\begin{figure}
\centerline{\epsfbox{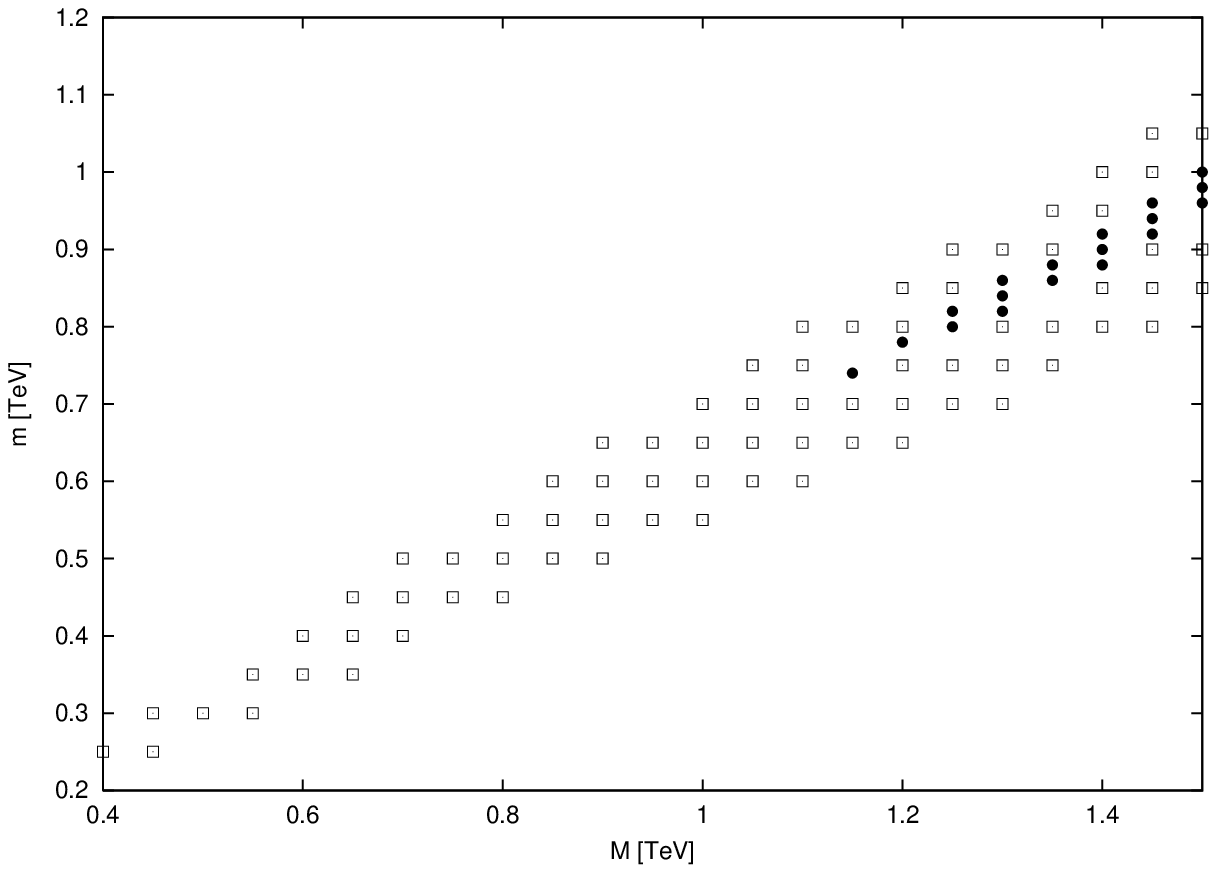}}
\caption{}
\label{fig1}
\end{figure}

\end{document}